\documentclass[10pt,a4paper]{article}
\usepackage[dvips]{color}
\usepackage{epsfig}
\usepackage{epstopdf}
\usepackage{amsmath}
\usepackage{graphicx}
\usepackage{authblk}
\usepackage{amssymb,amsmath}
\usepackage{amsmath}
\usepackage{tabularx}

\begin{document}
\title{\bf Phenomenological modification of horizon temperature}
\author{{M. Khurshudyan$^{a,b,c}$\thanks{Email:khurshudyan@yandex.ru, khurshudyan@tusur.ru}~~and As. Khurshudyan
$^{c}$\thanks{Email:khurshudyan@mechins.sci.am}}\\
$^{a}${\small {\em International Laboratory for Theoretical Cosmology, Tomsk State University of Control Systems and Radioelectronics (TUSUR), 634050 Tomsk, Russia}}\\ 
$^{b}${\small {\em Research Division,Tomsk State Pedagogical University, 634061 Tomsk, Russia}}\\
$^{c}${\small {\em Institute of Physics, University of Zielona Gora, Prof. Z. Szafrana 4a, 65-516 Zielona Gora,
Poland}}\\
$^{d}${\small {\em Institute of Mechanics, National academy of sciences (NAS) of Armenia, 24/2 Baghramyan ave., 0019 Yerevan, Armenia}}\\
}\maketitle

\begin{abstract}
In this paper a study of the accelerated expansion problem of the large scale universe is presented. To derive Friedmann like equations, describing the background dynamics of the recent universe we take into account, that it is possibile to interpret the spacetime dynamics as an emergent phenomenon. It is a consequence of the deep study of connection between gravitation and thermodynamics. The models considered are based on phenomenological modifications of the horizon temperature. In general, there are various reasons to modify the horizon temperature, one of which is related to the feedback from the spacetime on the horizon, generating additional heat. In order to constrain the parameters of the models we use $Om$ analysis and the constraints on this parameter at $z=0.0$, $z=0,57$ and $z=2.34$.
\end{abstract}

\section{Introduction}\label{sec:INT}

Possible solutions for a long standing problem, known as accelerated expansion of the large scale universe, have been discussed in literature very intensively, starting from the moment when it was detected by means of observational data~\cite{PC}~(and references therein). If we believe, that the general relativity describes the background evolution, the solution of the problem requires introduction of dark energy. On the other hand, to make the physics working, we need another dark component, known as dark matter. There are various interpretations of the dark energy, including scalar field and fluid interpretations. The interpretation in this case may be understood as a way to establish some functional connection between the pressure and the energy density of the component.

It is clear, that in the case of scalar field interpretation, we use a scalar field to define the pressure and the energy density of the dark energy. However, to have a complete description in this model of dark energy, we need to have the form of the potential, describing the scalar field. Unfortunately, at the moment it is not possible to determine the form of the potential from fundamental theories. Therefore, in literature various particular forms of the scalar field potential are considered. The scalar field interpretation is also very useful in the case of explaining the cosmic inflation, which is an accelerated expansion of the early universe. It is strongly believed, that for having an observable universe the required seeds are formed and originated exactly during the cosmic inflation. However, it does not completely exclude the cyclic nature of our universe, therefore the resulting picture can be inexact and should be modified with new observational data~\cite{MR}~(and references therein).

Different dark fluids are actively considered in literature. Usually they are described by the state equation, which provides a functional connection between the pressure and the energy density of the dark fluid. In recent literature, various parameterizations of the pressure, energy density and state equation parameter are studied according to the cosmological redshift, for instance, to describe the dark energy. In practice, the dark energy is introduced into the theory by hand and only after comparison of results with observational data it is possible to obtain some partial constraints (if we take into account existing tension between different datasets) on the models of dark energy~(see, for instance, \cite{F1}~-~\cite{F7}~ and references therein). On the other hand, a modification of the general relativity can provide appropriate source of negative pressure, characterizing the dark energy. In this case also we must consider that any modification should pass cosmological and astrophysical tests. From such perspective, as in the case of the dark energy, a big part of suggested modifications passess such tests~\cite{MG1}~-~\cite{MG12}. Comparing the two mentioned approaches, it is very hard to clearly indicate advantages of one approach over the other one so far.

In scientific literature, there is an increasing attention towards the models, where correctly chosen particle creation rate from dust, for instance, can accelerate the expansion of the universe. There are various models with phenomenologically chosen particle creation rates, that pass cosmological tests~\cite{PC1}~-~\cite{PC5}~(and references therein). However, it is clear, that this approach, as two previous approaches, represents particular examples of a more general theory, which should be developed. In relatively earlier works, authors express a hope, that with understanding of the structure of the dark energy, other problems will be solved and that to this end more observational data are needed. However, nowadays, an increasing amount of good observational data does not allow neither to prove, nor to reject this hypothesis, leaving central questions of modern cosmology and theoretical physics open.

In this paper we study the accelerated expansion of the large scale universe following to a very well-known idea by T. Padmanabhan~\cite{TP1}~-~\cite{TP3}. Namely, during the study we accept that the cosmic acceleration is observed, since the spacetime dynamics is an emergent phenomenon. Such breakthrough is directly related to the discovery of black hole thermodynamics, allowing better understanding of the nature of gravity. Deep study of the connection between gravity and thermodynamics reveals the emergent nature of gravity, i.e. gravity may not be a fundamental interaction. Argumentation of Padmanabhan, that the spatial expansion of our universe is due to the difference between the surface degrees of freedom~(DOF) and the bulk DOF in the region of emerged space allows to obtain equations, describing the background dynamics of the universe. Different manipulations with this idea, including modifications of bulk and surface DOFs can be find in literature, giving interesting alternative look to old results~\cite{EG1}~-~\cite{EG6}. In this paper we concentrate our attention on a possible modification of the horizon temperature, which can be due to feedback from the spacetime on the horizon.

In the next section we will demonstrate, that considered modifications provide appropriate Friedmann like equations, allowing to study the background dynamics. On the other hand, we perform $Om$ analysis to see possible departures from the $\Lambda$CDM for considered models, taking into account the fact, that for the standard model $Om = \Omega^{(0)}_{m}$ (the present day value of the dark matter)~\cite{Om}. We assume, that the effective fluid can be approximated as a fluid consisting of cold dark matter and a barotropic dark fluid with constant equation of state parameter.

The remainder of this paper describes in detail the models (section \ref{sec:MOD}), while in section \ref{sec:OC} the results from the cosmographic analysis are presented. In section \ref{sec:OMSHA} $Om$ and statefinder hierarchy analysis of the models are performed. In section \ref{sec:Discussion} the obtained results are summarized.

\section{Models}\label{sec:MOD}
In this section we start our discussion of the most simplest case, originally given by Padmanabhan, which states that the spatial expansion of our universe is due to the following general algorithm
\begin{equation}
\frac{dV}{dt} = L^{2}_{p}f( N_{sur} - N_{bulk}),
\end{equation} 
where $N_{sur}$ denotes the number of DOF on the spherical surface of Hubble radius $H^{-1}$ and reads as
\begin{equation}
N_{sur} = \frac{4 \pi H^{-2}}{L^{2}_{P}},
\end{equation}
with $L_{P}$ being the Planck length. On the other hand, the bulk DOF, $N_{bilk}$, reads as
\begin{equation}
N_{bulk} = \frac{|E|}{0.5 T},
\end{equation} 
where $|E| = |\rho + 3 P| V$ and $T$ is the horizon temperature.

In this paper, we will consider two phenomenological models for the horizon temperature. The first modification reads as
\begin{equation}\label{eq:T1}
T = \alpha  \frac{ H^{\beta }}{2 \pi },
\end{equation}
and the second one--
\begin{equation}\label{eq:T2}
T = \frac{H}{2\pi} + \alpha H^{\beta}.
\end{equation}
In both cases $\alpha$ and $\beta$ are constants and should be determined from the observational data. It is easy to see that the horizon temperature given by Eq.~(\ref{eq:T1}) will reduce to the usual horizon temperature with $\alpha = 1$ and $\beta = 1$, while the modified horizon temperature given by Eq.~(\ref{eq:T2}) with $\alpha = 0$ will reduce to $T = H/2\pi$. The cosmographical analysis presented in the next section will be extended using $Om$ analysis-- a geometrical tool to study dark energy models involving the following parameter~\cite{Om}
\begin{equation}
Om = \frac{x^{2}-1}{(1+z)^{3} - 1},
\end{equation}  
where $x = H/H_{0}$ and $H_{0}$ is the value of the Hubble parameter at $z=0$. Note, that the $Om$ analysis is generalized to the two point $Om$ analysis with~\cite{Om2}
\begin{equation}
1Om(z_{2},z_{1}) = \frac{x(z_{2})^{2} - x(z_{1}^{2})}{(1+z_{2})^{2} - (1+z_{1})^{2}}.
\end{equation}
Moreover, a slight modification of the two point $Om$ is suggested ($Omh^{2}$) in Ref.~\cite{Om2} and the estimated values of the two point $Omh^{2}$ for $z_{1} = 0$, $z_{2} = 0.57$ and $z_{3} = 2.34$:
$$Omh^{2}(z_{1};z_{2}) = 0.124 \pm 0.045,$$
$$Omh^{2}(z_{1};z_{3}) = 0.122 \pm 0.01,$$
\begin{equation}
Omh^{2}(z_{2};z_{3}) = 0.122 \pm 0.012,
\end{equation}
will be used to obtain constraints on the parameters of the models. Recall, that for the $\Lambda$CDM model the value of $Omh^{2}=0.1426$.

\section{Cosmography}\label{sec:OC}
To simplify the discussion we separate the material into two subsections. Results obtained in this section will be used in the next section to complete the study of the models with the $Om$ and statefinder hierarchy analysis. 

\subsection{Model 1}
The consideration of the horizon temperature given by Eq.~(\ref{eq:T1}) gives a model of the universe, where the dynamics of the Hubble parameter reads as 
\begin{equation}\label{eq:dH1}
\dot{H} + H^{2} = \frac{4 \pi  G}{3} \frac{ H^{1-\beta} |\rho + 3 P | }{\alpha}.
\end{equation}
The last equation can be integrated in quadratures easily:
\begin{equation}
H(z) = \left [ C_1 (1+z)^{1 + \beta} + \frac{  \rho_{0} (1 + \beta) (1 + 3 \omega) (1+z)^{3 (1 + \omega)}}{6 \alpha  (-\beta +3 \omega +2)} \right ]^{\frac{1}{1+\beta}},
\end{equation}
describing the accelerated expanding phase with $\rho+3P < 0$. Here $C_1$ is the integration constant and can be determined from the Cauchy condition $H(0) = H_{0}$. During the integration we take into account, that the dynamics of the effective fluid, described by $P$ and $\rho$ reads as
\begin{equation}
-(1+z)\frac{d\rho}{dz} + 3 \rho (1+ \omega ) = 0.
\end{equation}
On the other hand, the integration of Eq.~(\ref{eq:dH1}) with $\rho+3P > 0$ gives the model of the universe with the following Hubble parameter
\begin{equation}
H(z) = \left [C_1 (1+z)^{1 + \beta} - \frac{\rho_{0} (1 + \beta) (1 + 3 \omega ) (1+z)^{3 (1+ \omega)}}{6 \alpha  (-\beta +3 \omega +2)} \right]^{\frac{1}{1 + \beta}}
\end{equation}

\begin{figure}[h!]
 \begin{center}$
 \begin{array}{cccc}
\includegraphics[width=80 mm]{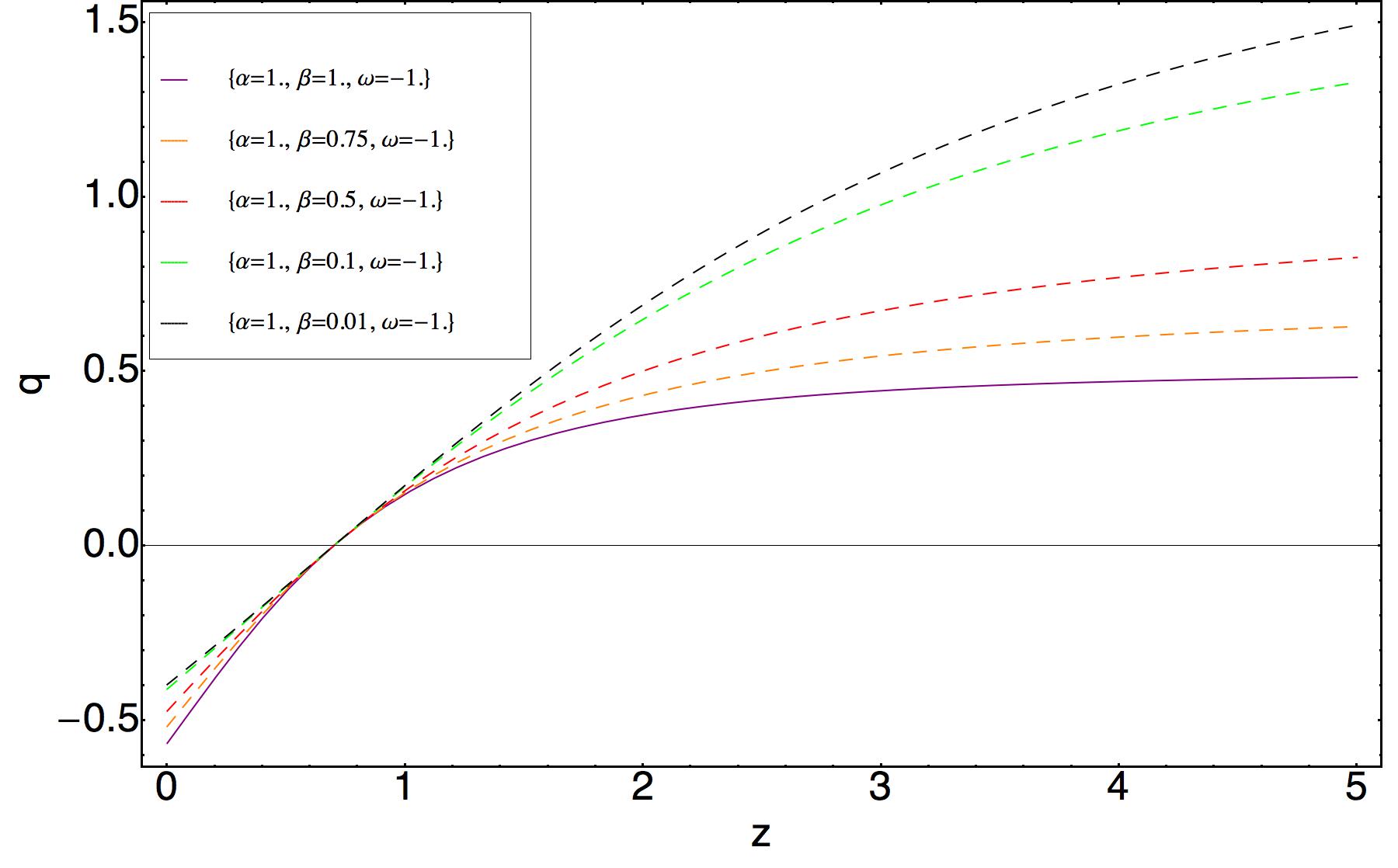}  &
\includegraphics[width=80 mm]{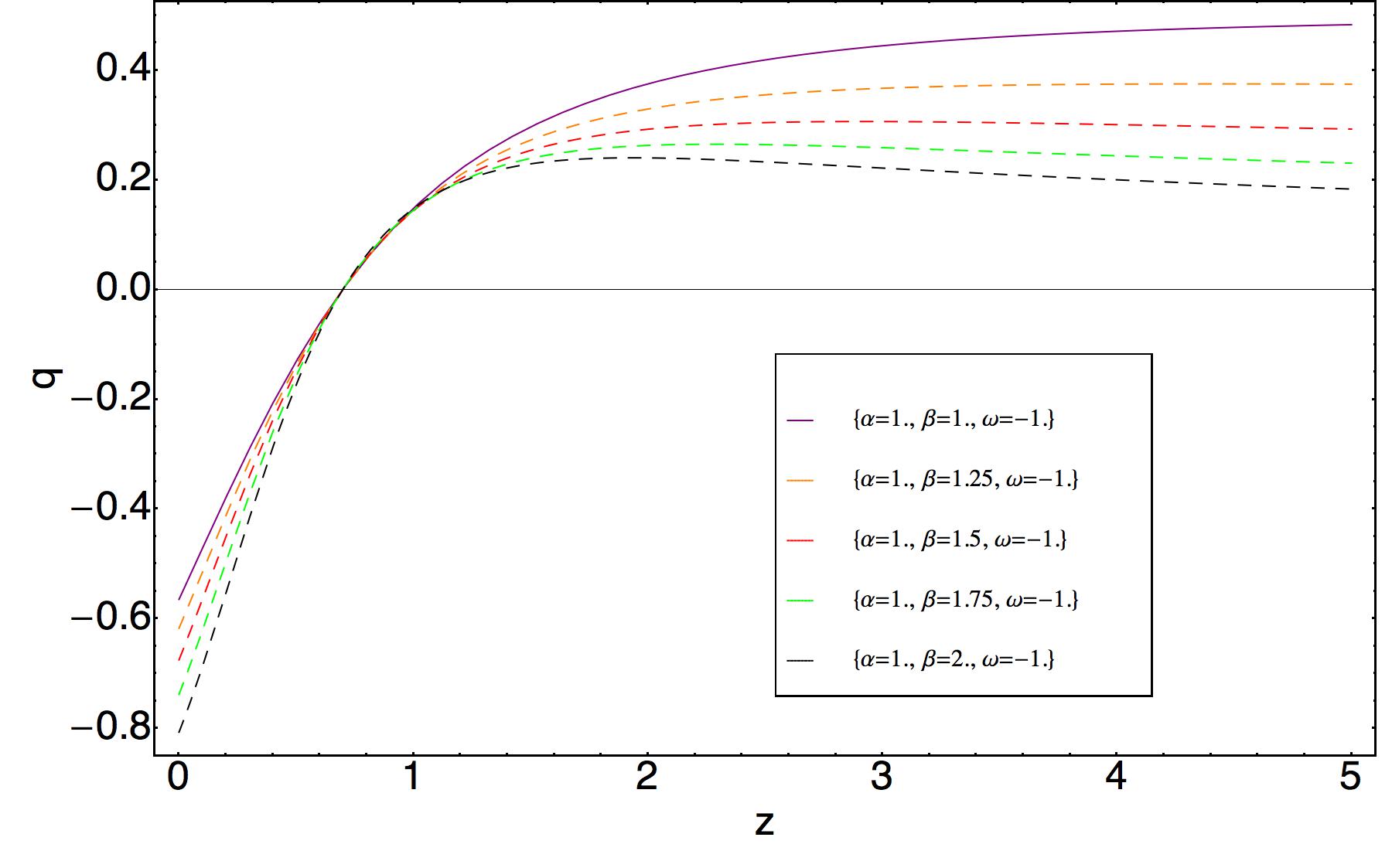}  \\
 \end{array}$
 \end{center}
\caption{Graphical behavior of the deceleration parameter $q$ for the universe with two components fluid. The horizon temperature is given by Eq.~(\ref{eq:T1})}
 \label{fig:Fig1}
\end{figure}

In what follows, to study the accelerated expansion of the large scale universe, we assume, that the effective fluid consists of two components: cold dark matter and dark energy, described by barotropic fluid equation with constant (negative) equation of state parameter. To see possible departures form the $\Lambda$CDM standard model, we take $\alpha = 1$ and $\omega = -1$. The graphical behavior of the deceleration parameter $q$ for the universe with two components fluid described above is presented in Fig.~(\ref{fig:Fig1}). In both plots the blue line with $\beta = 1$ corresponds to the $\Lambda$CDM model with $\Omega^{(0)}_{dm} = 0.29$ and $H_{0} = 0.7$. The left plot represents the behavior of the deceleration parameter for $0 < \beta \leq 1$. The second plot represents the behavior of the same parameter for $1\leq \beta \leq 2$.

In both cases, a phase transition between universe with decelerated expansion into that with accelerated expansion is evident. Moreover, the transition redshift does not depend on the parameter $\beta$. However, higher and lover redshift behaviors depend on the parameter $\beta$. In particular, in the $0 < \beta \leq 1$ case, a decrease of $\beta$ will increase $q$ compared to $q_{\Lambda CDM}$ for higher redshifts (in a decelerated expanding phase). It is also the case for lower redshifts, when accelerated expanding of the universe occurs.

\begin{figure}[h!]
 \begin{center}$
 \begin{array}{cccc}
\includegraphics[width=80 mm]{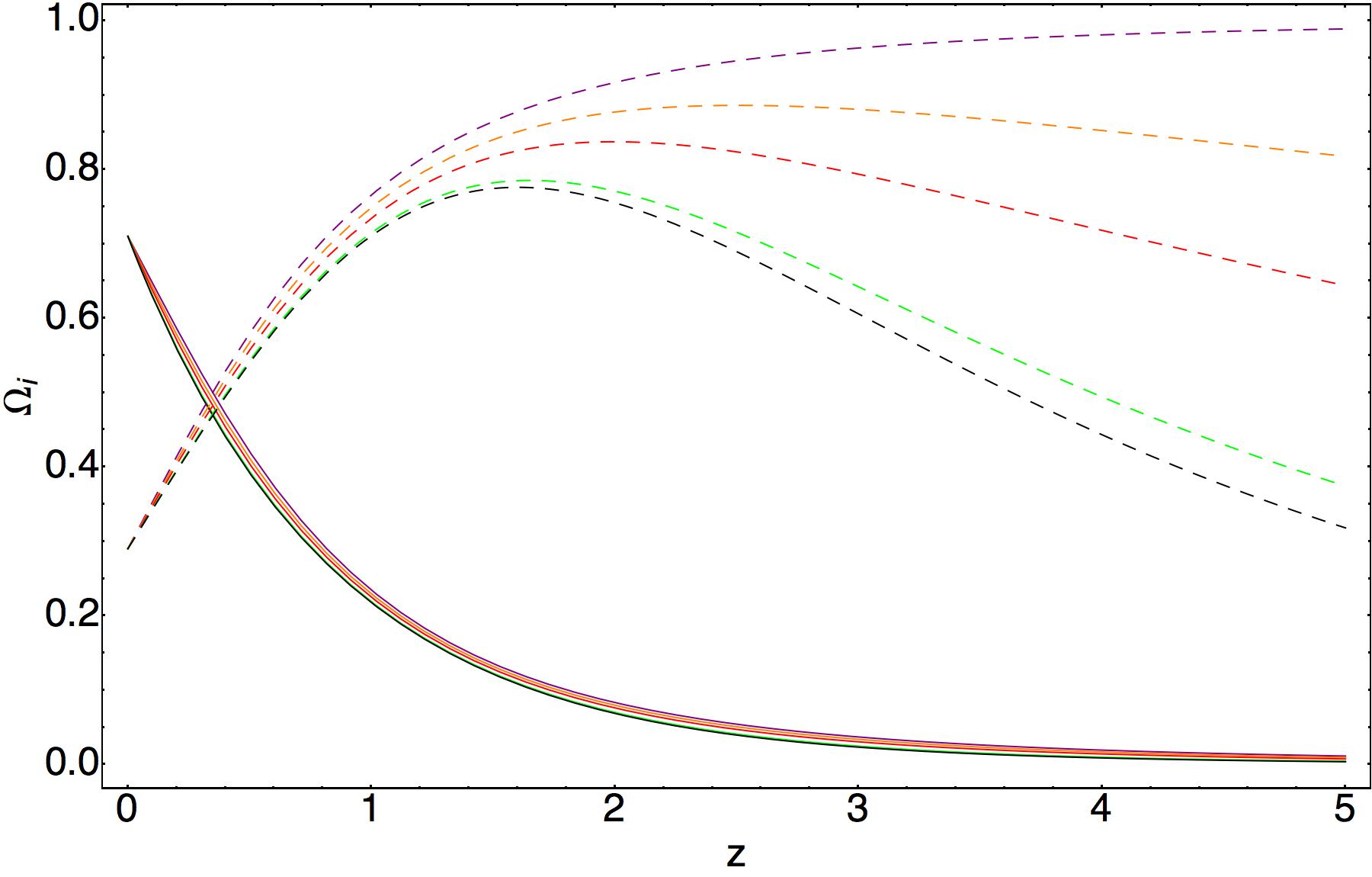}  &
\includegraphics[width=80 mm]{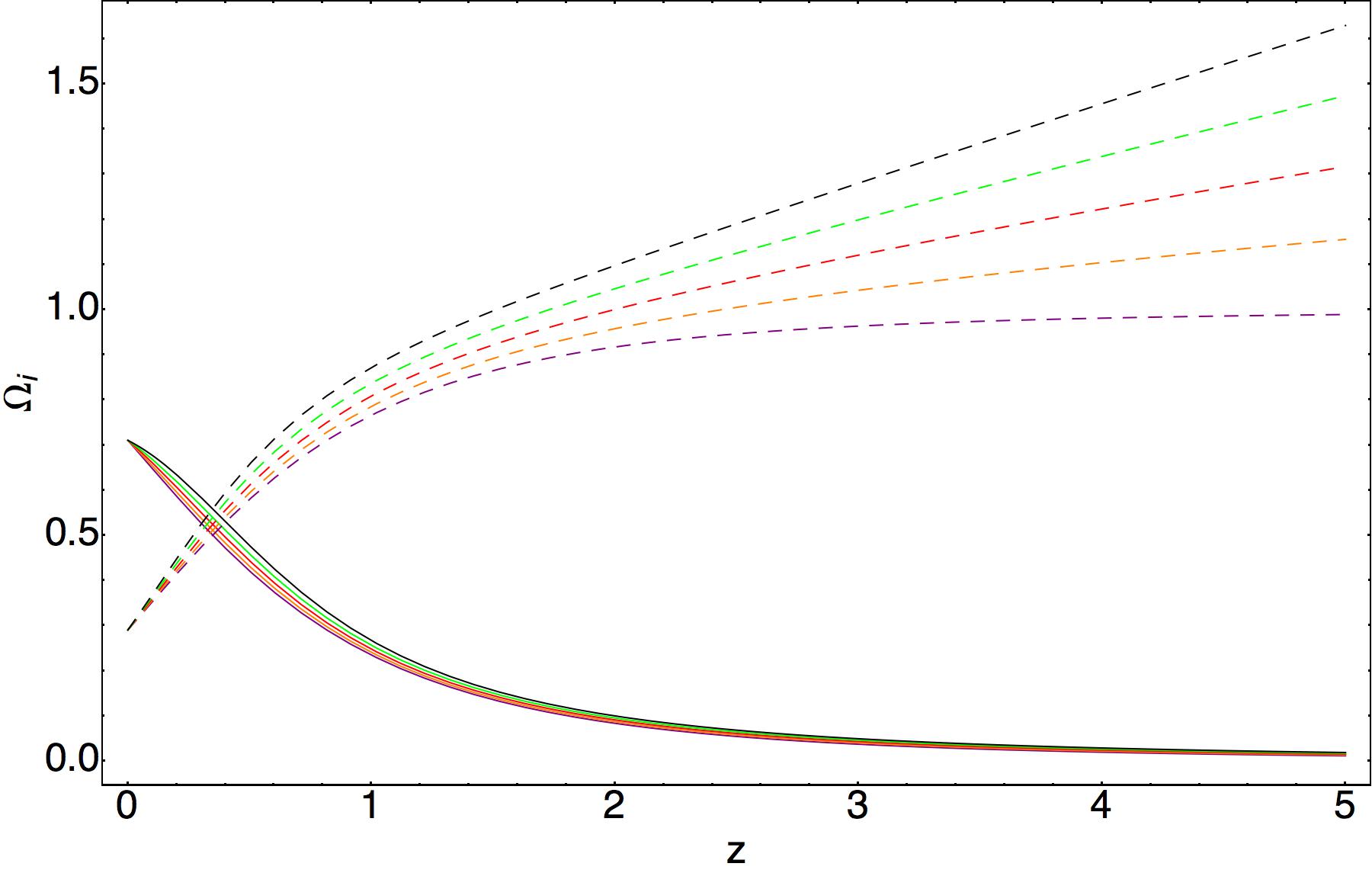}  \\
 \end{array}$
 \end{center}
\caption{Graphical behavior of $\Omega_{de}$~(solid lines) and $\Omega_{dm}$~(dashed lines) parameters with $\Omega_{de} + \Omega_{dm} = 1$ for $z=0$. The horizon temperature is given by Eq.~(\ref{eq:T1})}
 \label{fig:Fig2}
\end{figure}

Opposite impact of the parameter $\beta$ on $q$ is observed in the second case, described by $1\leq \beta \leq 2$. The graphical behavior of $\Omega_{de}$ and $\Omega_{dm}$ are presented in Fig.~(\ref{fig:Fig2}) with the present value of $\Omega_{de} + \Omega_{dm}$ normalized to $1$. It can be seen, that for higher redshifts when $0 < \beta \leq 1$, we can have models of open universe appointed just by the amount of the cold dark matter in the universe. As was expected, with $\beta = 1$ the model becomes $\Lambda$CDM model, where the cosmological constant problems are solved. On the other hand, for the case with $1\leq \beta \leq 2$, we clearly see, that cold dark matter can generate a model of a "closed" universe just due to the amount of cold dark matter. However, independently from the feature of the model of the universe for higher redshift, the recent universe is a flat universe.

\subsection{Model 2}
\begin{figure}[h!]
 \begin{center}$
 \begin{array}{cccc}
\includegraphics[width=80 mm]{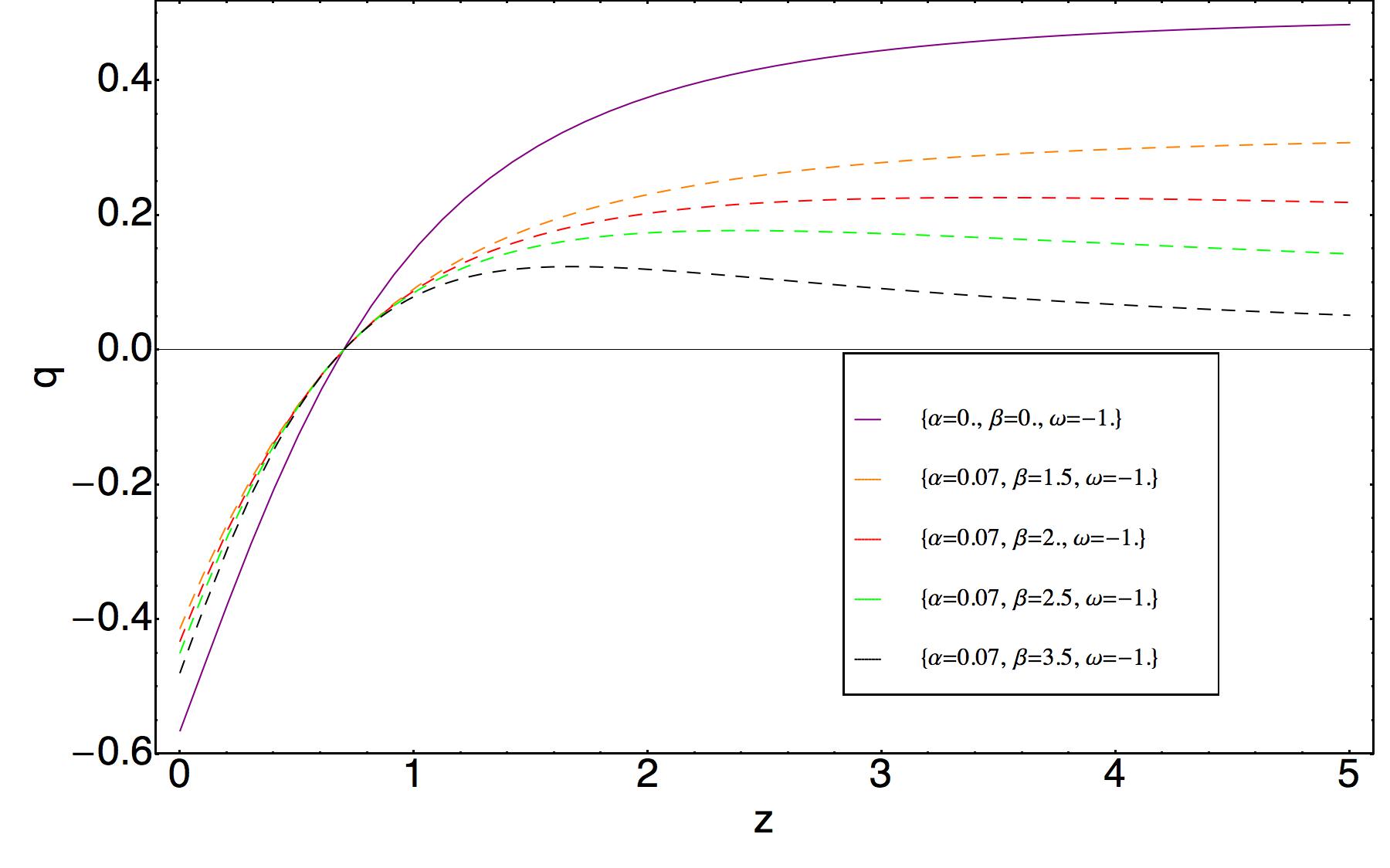}  &
\includegraphics[width=80 mm]{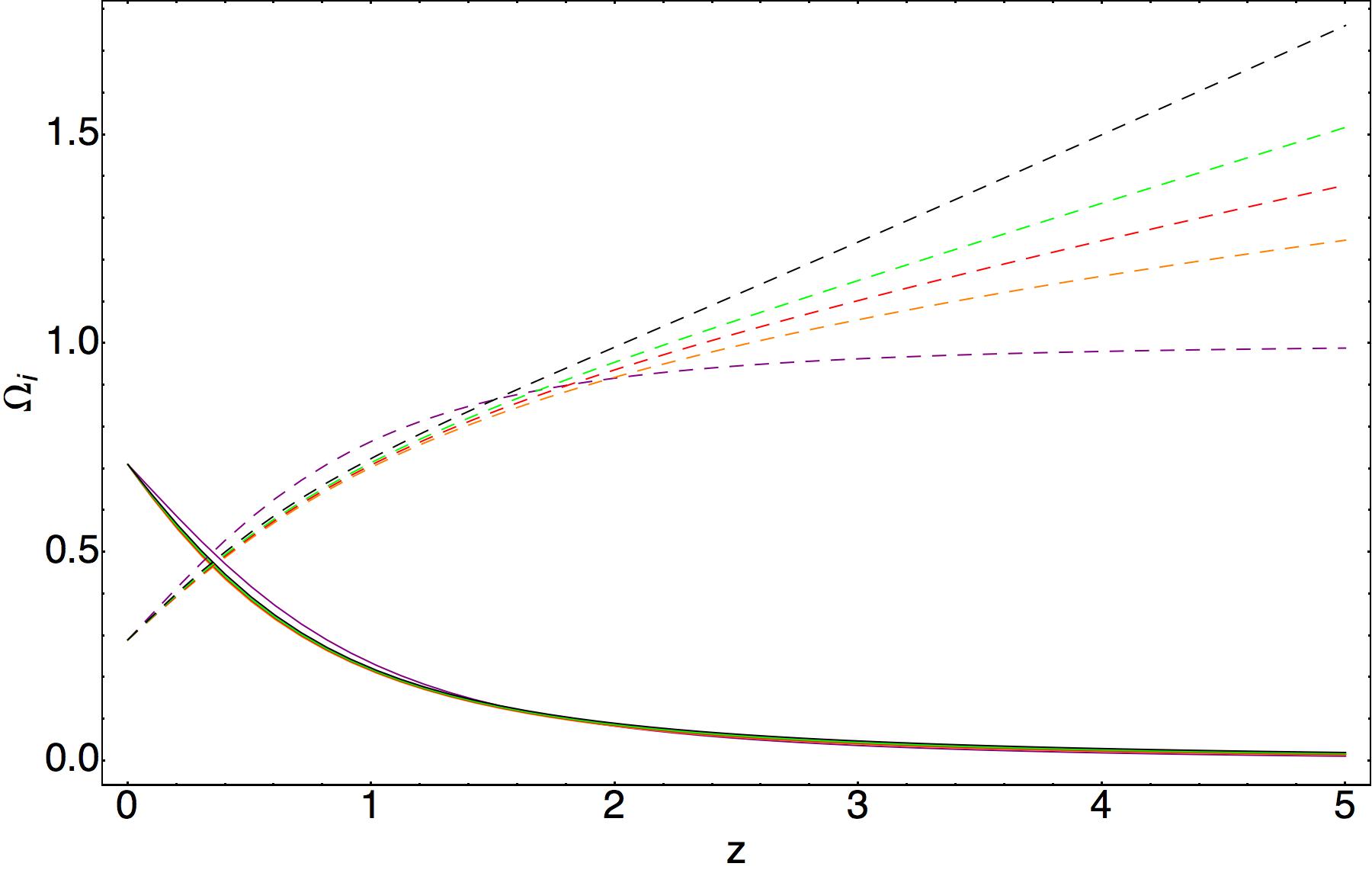}  \\
 \end{array}$
 \end{center}
\caption{Graphical behavior of the deceleration parameter $q$ for the universe with two component fluid is presented on the left plot. The graphical behavior of $\Omega_{de}$~(solid lines) and $\Omega_{dm}$~(dashed lines) parameters with $\Omega_{de} + \Omega_{dm} = 1$ for $z=0$  the same model is presented on the right plot. The horizon temperature is given by Eq.~(\ref{eq:T2})}
\label{fig:Fig3}
\end{figure}

The second model considered in this section is described by the horizon temperature given by Eq.~(\ref{eq:T2}). Such considerations give a model of the universe with the following dynamics for the Hubble parameter
\begin{equation}\label{eq:dH2}
\dot{H} + H^{2} =  \frac{4 \pi  G}{3} \frac{| \rho + 3 P |}{(1 + 2 \pi \alpha H^{\beta -1 })}.
\end{equation}
Integration of the last equation with an assumption $\rho+3P < 0$ for $\beta = 1$~(a particular solution) gives the following dynamics for the Hubble parameter 
\begin{equation}
H(z) = \frac{(z+1) \sqrt{3 C_2 (2 \pi  \alpha +1)  + \rho_{0}(z+1)^{3 \omega +1}}}{\sqrt{6 \pi  \alpha +3}},
\end{equation}
where $\rho_{0}$ is the energy density value of the effective fluid with $P = \omega \rho$ for $z = 0$. The integration constant $C_2$ will be determined from the Cauchy condition $H(0) = H_{0}$. Following to the two interpretation of the effective fluid for the universe with the horizon temperature given by Eq.~(\ref{eq:T2}), we study the graphical behavior of the deceleration parameter numerically. In this case we compare the new model with the standard $\Lambda$CDM model. The comparison shows, that the transition redshift is not affected and for the higher redshits an increase of the parameter $\beta$ causes decrease of the deceleration parameter. It is seen from the graphical behavior presented in Fig.~(\ref{fig:Fig3}), that for higher redshift we decelerate expanding universe. On the other hand, at low redshifts, the same increase of $\beta$ increases the decelerated parameter. Moreover, the numerical analysis of $\Omega_{de}$ and $\Omega_{dm}$ indicates, that for higher redshifts a closed universe is "observed" due to "curvature", generated by the cold dark matter. The presented graphical behaviors of all three parameters are due to the constrains from $Om$ analysis, which is performed in the next section. 

\section{$Om$ and statefinder hierarchy analysis}\label{sec:OMSHA}

\begin{figure}[h!]
 \begin{center}$
 \begin{array}{cccc}
\includegraphics[width=80 mm]{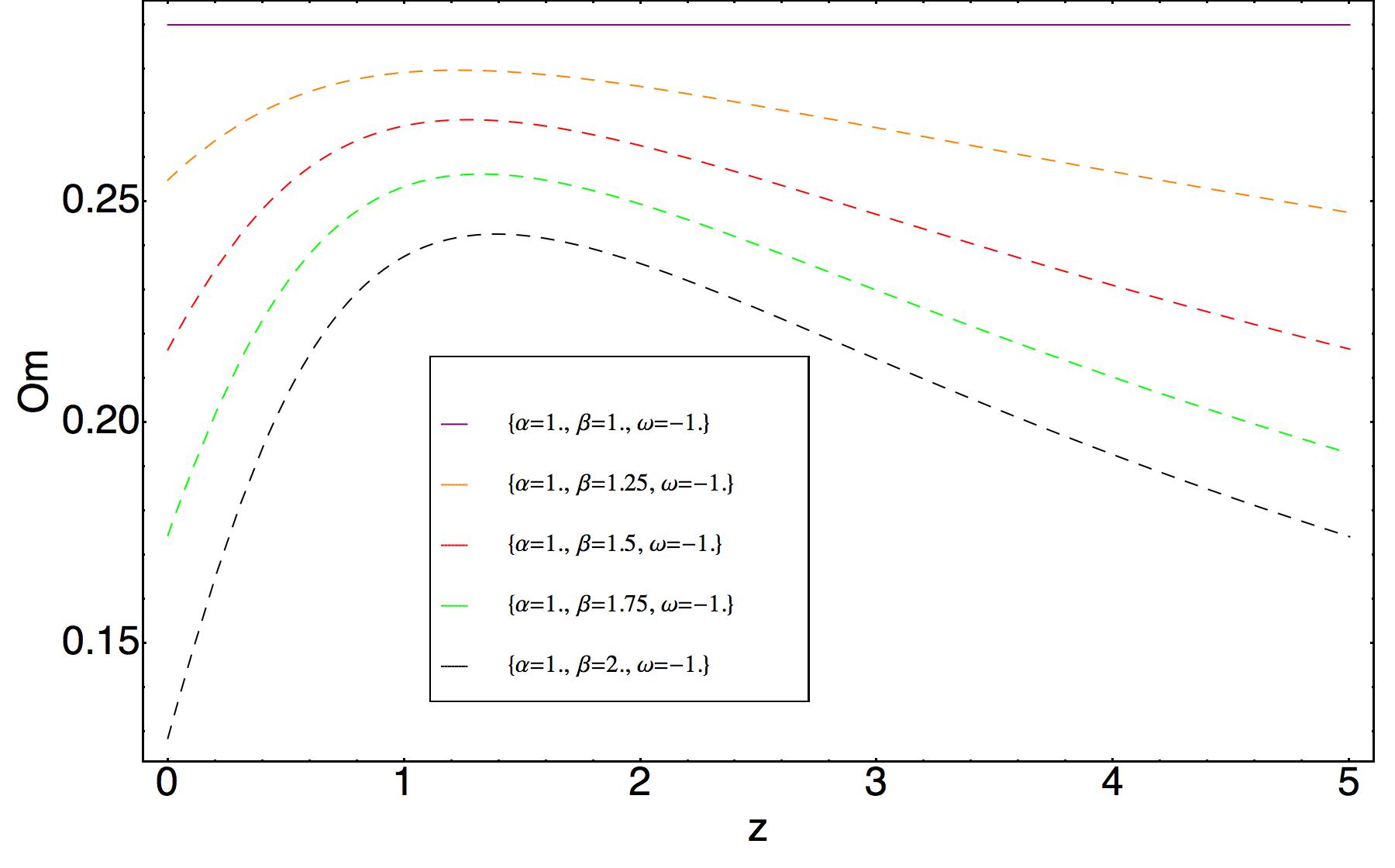}  &
\includegraphics[width=80 mm]{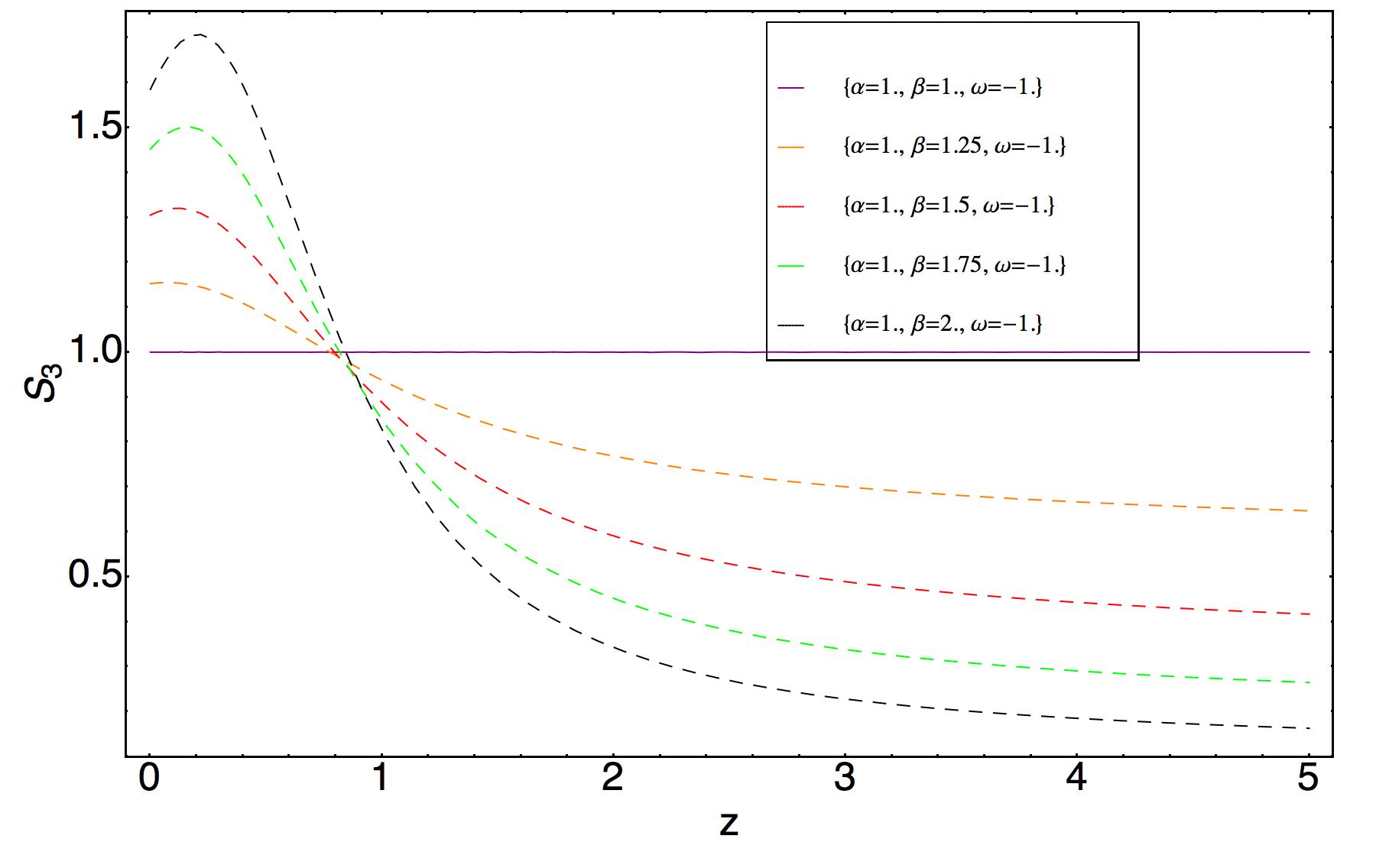}  \\
 \end{array}$
 \end{center}
\caption{Graphical behavior of the $Om$ and the $S_{3}$ parameters for the model with the horizon temperature is given by Eq.~(\ref{eq:T1}). $\Omega_{de} + \Omega_{dm}$ is normalized to $1$ at $z=0$}
\label{fig:Fig4}
\end{figure}

As it is mentioned above, $Om$ analysis is one of the basic tools to study dark energy models. On the other hand, a modification of it can be used to obtain some constraints on the parameters of the models. Another tool to analyze dark energy models is the so-called statefinder hierarchy analysis with the following parameters~\cite{SH}
\begin{equation}\label{eq:S3}
S^{(1)}_{3} = A_{3},
\end{equation}
\begin{equation}\label{eq:S4}
S^{(1)}_{4} = A_{4} + 3(1+q),
\end{equation}
\begin{equation}\label{eq:S5}
S^{(1)}_{5} = A_{5}  - 2 (4+ 3q)(1+q),
\end{equation}
etc., where $q$ is the deceleration parameter, while $A_{n}$ reads as
\begin{equation}
A_{n} = \frac{a^{(n)}}{a H^{n}},
\end{equation} 
with 
\begin{equation}
a^{(n)} = \frac{d^{n}a}{dt^{n}}.
\end{equation}
Statefinder hierarchy for the $\Lambda$CDM model when the cosmic expansion is equal to $1$. However, for models with a dynamical dark energy and dark matter, $S^{(1)}_{n}$ are varying quantities and the $\Lambda$CDM model can be chosen as a reference frame to emphasize possible deviations. For instance, the behavior of $Om$ parameter with its increasing and decreasing behavior indicates, that the first model with two component fluid is different from the $\Lambda$CDM model~(the blue line on the left plot of Fig.~(\ref{fig:Fig4}) indicates the $\Lambda$CDM).

\begin{figure}[h!]
 \begin{center}$
 \begin{array}{cccc}
\includegraphics[width=80 mm]{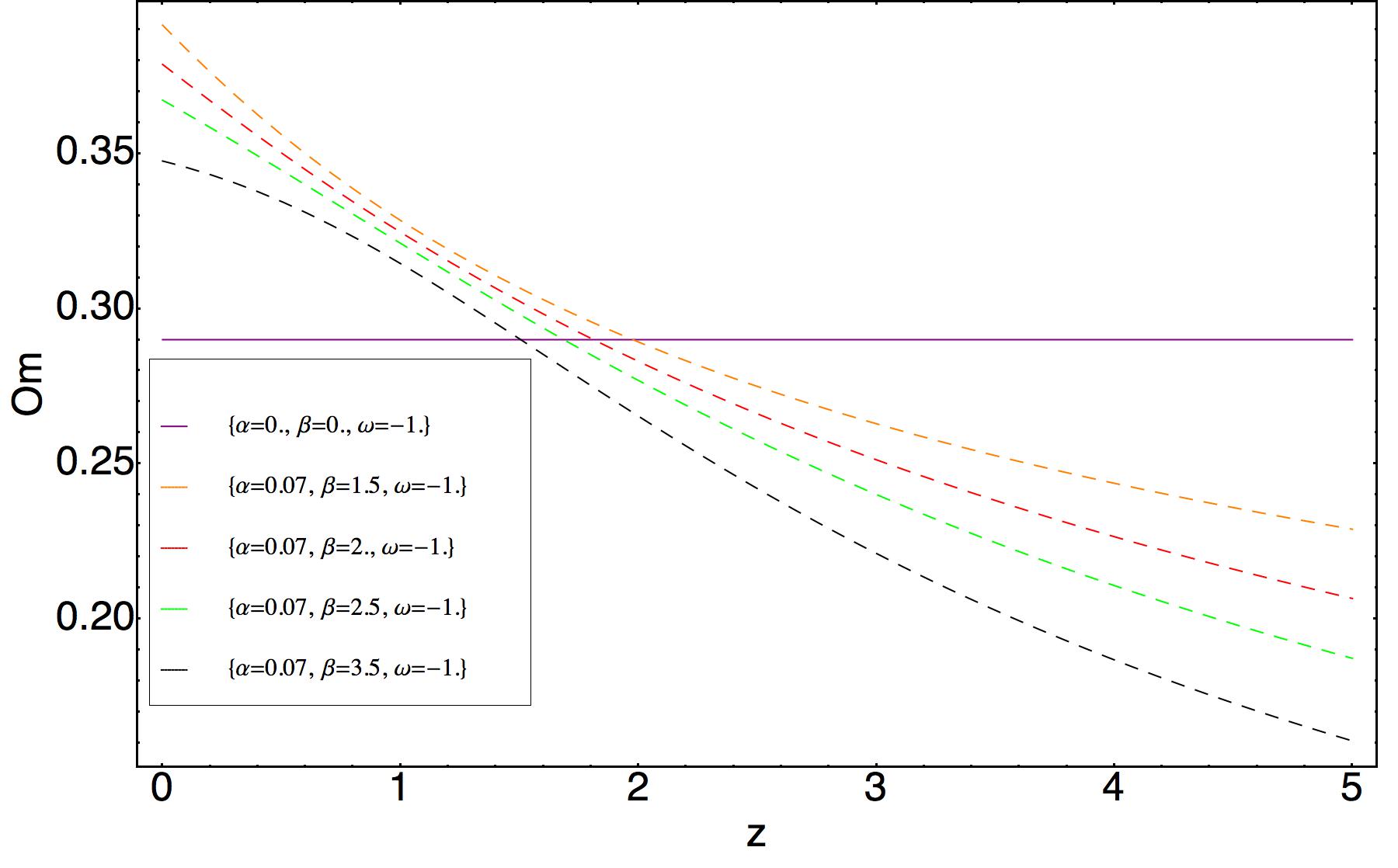}  &
\includegraphics[width=80 mm]{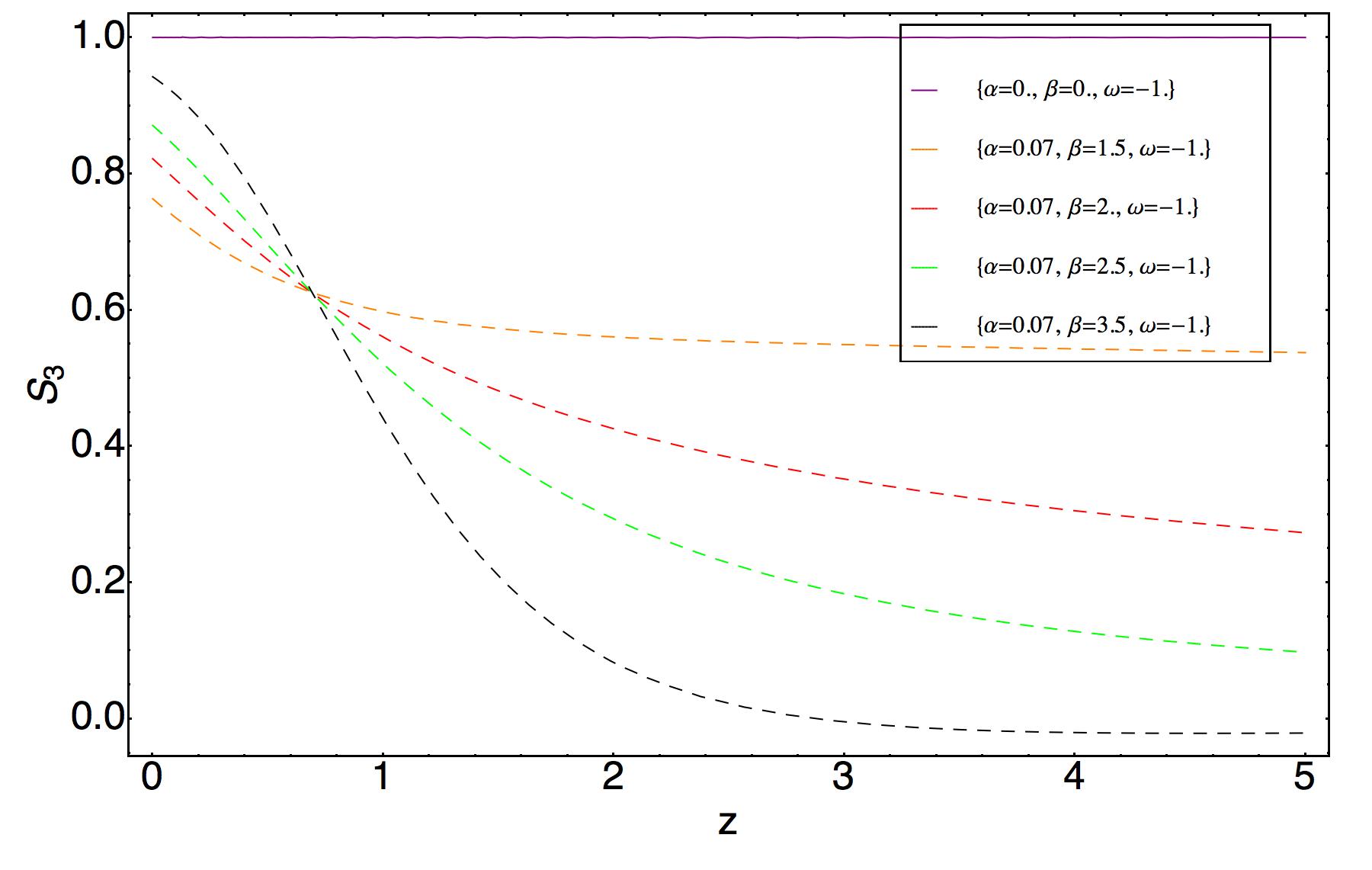}  \\
 \end{array}$
 \end{center}
\caption{Graphical behavior of the $Om$ and the $S_{3}$ parameters for the model when the horizon temperature is given by Eq.~(\ref{eq:T2}). $\Omega_{de} + \Omega_{dm}$ is normalized to $1$ at $z=0$}
 \label{fig:Fig5}
\end{figure}

The study of the $Om$ parameter does not indicate a possibility of the new model to be the same with the $\Lambda$CDM standard model. However, for instance, the study of $S_{3}$ parameter from the statefinder hierarchy analysis shows when~(at which redshifts) the model becomes $\Lambda$CDM. On the other hand, for the second model with the horizon temperature given by Eq.~(\ref{eq:T2}), $Om$ analysis will indicate when~(at which redshifts) the model will become $\Lambda$CDM, but $S_{3}$ parameter can not do it~(Fig.~(\ref{fig:Fig5})). However, both parameters for both models can clearly demonstrate features and differences of the models. The values of the parameters of the models are chosen to satisfy the constraints from the $Om$ analysis discussed in section~\ref{sec:MOD} to reduce our discussion.

\section{Discussion}\label{sec:Discussion}

In this paper two phenomenological modifications of the horizon temperature are considered. The spacetime dynamics is accounted to be an emergent phenomenon. Following to the general receipt, Friedmann like equation for the dynamics of the Hubble parameter are obtained for both models. On the other hand, we assume, that the effective fluid for the models can be approximated as a two component fluid with cold dark matter and dark energy. Moreover, the dark energy is assumed to be represented as a barotropic fluid with negative equation of state parameter. After such assumptions, the cosmological model becomes interesting for us to compare with the $\Lambda$CDM standard model when $H_{0}=0.7$, $\Omega^{(0)}_{m} = 0.29$ and $\omega = -1$.

To make the comparison more realistic, we used constraints obtained form the $Om$ analysis known from the literature. In the case of the first model, two ranges for the parameter $\beta$ are considered and it turns out, that the cosmographic analysis with $0 < \beta \leq 1$ shows a reasonable behavior for the cosmological parameters. The similar picture is observed with $1 \leq \beta \leq 2$. However, only taking into account above mentioned constraints allows us to observe, that the model with $1 \leq \beta < 2.5$ is a favorable model.

A similar analysis also allows us to obtain an appropriate range for the parameter $\beta$ for the second model: $\alpha = 0.07$ and $2 \leq \beta < 4.5$. With such constraints we see, that for both models, an observer at higher redshifts due to the amount of the cold dark matter can interpret such universes to be "closed". At the same time, both models at low redshifts can effectively explain the accelerated expansion of the universe. Moreover, as we can see from the graphical behaviors of the cosmological parameters, the models are free from the problems associated with the cosmological constant problem.

In other words, our analysis shows, that a simple model of dark energy (cosmological constant) with appropriate modification of the horizon temperature can be used to solve the three main problems in the modern cosmology. On the other hand, $Om$ and statefinder hierarchy analysis shows, that both tools are very good applicable for suggested models. Moreover, one of the tools shows features, which can not be identified with other tools. Therefore, they can be used together to fulfill the results obtained by one of them by the other one. The constraints on the $\beta$ and $\alpha$ parameters from different datasests using $\chi^{2}$ statistical technique, should be used in order to narrow the constraints obtained from an extended $Om$ analysis. It will be also interesting to study the problems of structure formation. However, such tasks are left as a subject of further considerations.   

\section*{\large{Acknowledgments}}
The authors appreciate Prof. K. Urbanowski from Institute of Physics, University of Zielona Gora, for valuable comments/suggestions and for inviting their attention on Ref.-s~\cite{U1}~-~\cite{U6} during the preparation of the paper.

\end{document}